\documentclass[twocolumn,floatfix,superscriptaddress,a4paper,showpacs,showkeys,nofootinbib,reprint,prl]{revtex4-1}
\usepackage{epsfig}
\usepackage{latexsym}
\usepackage{xspace}
\usepackage[colorlinks=true,linktocpage=true,linkcolor=blue,citecolor=blue,allcolors=blue]{hyperref}
\usepackage{url}
\usepackage[utf8]{inputenc}
\usepackage{indentfirst}
\usepackage{enumerate}
\usepackage{color}

\usepackage[caption=false,position=top]{subfig}

\usepackage{amsmath}
\usepackage{amssymb}
\usepackage[english]{babel}
\usepackage{url}

\AtBeginDocument{\renewcommand{\natexlab}[1]{#1}}

\newcommand{\eq}[1]{\begin{align} #1 \end{align}}

\begin{document}


\title{
Pion Condensation in the Early Universe at Nonvanishing Lepton Flavor Asymmetry\\ 
and Its Gravitational Wave Signatures
}

\author{Volodymyr Vovchenko}
\affiliation{Nuclear Science Division, Lawrence Berkeley National Laboratory, 1 Cyclotron Road, Berkeley, CA 94720, USA}

\author{Bastian~B.~Brandt}
\affiliation{
Fakult\"at f\"ur Physik, Universit\"at Bielefeld, D-33615 Bielefeld, Germany.
}

\author{Francesca Cuteri}
\affiliation{
Institut f\"ur Theoretische Physik,
Goethe Universit\"at Frankfurt, Max-von-Laue-Str. 1, D-60438 Frankfurt am Main, Germany}

\author{Gergely Endr\H{o}di}
\affiliation{
Fakult\"at f\"ur Physik, Universit\"at Bielefeld, D-33615 Bielefeld, Germany.
}

\author{Fazlollah Hajkarim}
\affiliation{
Institut f\"ur Theoretische Physik,
Goethe Universit\"at Frankfurt, Max-von-Laue-Str. 1, D-60438 Frankfurt am Main, Germany}
\affiliation{
Dipartimento di Fisica e Astronomia, Universit\`a degli Studi di Padova, Via Marzolo 8, 35131 Padova, Italy}

\author{J\"urgen Schaffner-Bielich}
\affiliation{
Institut f\"ur Theoretische Physik,
Goethe Universit\"at Frankfurt, Max-von-Laue-Str. 1, D-60438 Frankfurt am Main, Germany}

\begin{abstract}
We investigate the possible formation of a Bose-Einstein condensed phase of pions in the early Universe at nonvanishing values of lepton flavor asymmetries.
A hadron resonance gas model with pion interactions, based on 
first-principle lattice QCD simulations at nonzero isospin density, is used
to evaluate cosmic trajectories at various values of electron, muon, and tau lepton asymmetries that satisfy the available constraints on the total lepton asymmetry.
The cosmic trajectory can pass through the pion condensed phase if the combined electron and muon asymmetry is sufficiently large: $|l_e + l_{\mu}| \gtrsim 0.1$,
with little sensitivity to the difference $l_e - l_\mu$ between the individual flavor asymmetries.
Future constraints on the values of the individual lepton flavor asymmetries will thus be able to either confirm or rule out the condensation of pions during the cosmic QCD epoch.
We demonstrate that the pion condensed phase leaves an imprint both on the spectrum of primordial gravitational waves
and on the mass distribution of primordial black holes at the QCD scale e.g. the black hole binary of recent LIGO event GW190521 can be formed in that phase. 
\end{abstract}



\maketitle


\paragraph*{Introduction.}

The origin of matter-antimatter asymmetry in the Universe is unknown as yet. There are several theoretical attempts to explain this fact which has to originate from the evolution of the very early Universe \cite{Riotto:1999yt,Cline:2006ts}. 
The asymmetry can be expressed in terms of the values of charges that are conserved in the standard model: baryon number $B$, electric charge $Q$ and lepton number $L$.
These numbers are conserved during the cosmic evolution following baryo- and lepto-genesis \cite{Riotto:1999yt,Cline:2006ts,Davidson:2008bu,Flanz:1994yx}. 
Neutrino oscillations start to occur in the early Universe at $T \sim 10$~MeV,
therefore, at higher temperatures not only the total lepton asymmetry is conserved, but also the individual electron, muon, and tau lepton asymmetries.
Conservation of these numbers leads to the evolution of chemical potentials of different particles that were present in the thermal bath and contributed to the equation of state of the Universe at early eras. 

Recently, the LIGO experiment detected several gravitational wave (GW) events from the merger of black holes predicted by general relativity  \cite{Abbott:2016blz,TheLIGOScientific:2016htt}.
GWs may also have a cosmic origin due to inflation or possible cosmic (phase) transitions 
\cite{Caprini:2018mtu}. Primordial gravitational waves (PGWs) can be produced from the perturbation of spacetime \cite{grishchuk1974amplification,Starobinsky:1979ty} by the inflationary phase in the early Universe \cite{Mukhanov:1990me}. Passing through the different stages of cosmic history like the QCD and electroweak transitions, and the matter dominated epoch will leave imprints on PGWs due to the variation of the Hubble expansion rate \cite{Watanabe:2006qe,Bernal:2019lpc,Hajkarim:2019csy,Schettler:2010dp}.

Black holes (BHs) can either form by the collapse of matter in stars or in the early Universe due to primordial density perturbations generated by inflation \cite{Carr:1975qj,Carr:1974nx}. 
The latter ones are known as primordial black holes (PBHs) -- possible dark matter candidates~\cite{Bird:2016dcv}.
The formation of PBHs is caused by the collapse of inhomogeneous high density regions during the time the modes cross the horizon~\cite{Khlopov:2008qy,Sasaki:2018dmp,Carr:2016drx}.
These processes depend on the inflationary scenario and the scales adopted, as well as on the thermal history of the early Universe, making them sensitive to the matter-antimatter asymmetry.

For an isentropic expansion of the Universe it is common to express the asymmetries in terms of the conserved charge per entropy ratios: $b = n_B/s$, $q = n_Q/s$, and $l_\alpha = n_{L_\alpha}/s$ with $\alpha = e,\mu,\tau$. 
One can associate a chemical potential to each of the conserved charges $B$, $Q$, and $\{L_\alpha\}$.
The cosmic trajectory is a line in the six-dimensional space of $T$, $\mu_B$, $\mu_Q$, $\mu_e$, $\mu_\mu$, and $\mu_\tau$
defined by five conservation equations:
\begin{align}
\label{eq:B}
\frac{n_B(T,\mu_B,\mu_Q)}{s(T,\mu_B,\mu_Q,\{\mu_\alpha\})} & = b,\\
\label{eq:Q}
\frac{n_Q(T,\mu_B,\mu_Q,\{\mu_\alpha\})}{s(T,\mu_B,\mu_Q,\{\mu_\alpha\})} & = 0,\\
\label{eq:l}
\frac{n_{L_\alpha}(T,\mu_Q,\{\mu_\alpha\})}{s(T,\mu_B,\mu_Q,\{\mu_\alpha\})} & = l_\alpha, \quad \alpha  \in e,\,\mu,\,\tau.
\end{align}
The conserved charge and entropy densities entering the above equations are given as functions of the temperature and chemical potentials through the equation of state of cosmic matter.
For the cosmic QCD epoch, the equation of state is mainly determined by  strongly interacting matter, but also contains the contributions of leptons and photons.
Naturally, non-trivial dynamics is mainly contained in the QCD part.

Tight constraints on the baryon asymmetry and electric charge are available: $b = (8.60 \pm 0.06) \times 10^{-11}$ and $q = 0$.
The total lepton asymmetry in the standard scenario arises through sphaleron processes, giving $l = -(51/28)b$, equally distributed among the three lepton flavors \cite{Harvey:1990qw}.
This yields the standard cosmic trajectory where all chemical potentials are close to vanishing for the majority of the cosmic trajectory.
Values of the total lepton asymmetry considerably larger than the baryon one are also possible: Ref.~\cite{Oldengott:2017tzj} gives the constraint of $|l| < 0.012$. 
Here $l = l_e + l_\mu + l_\tau$.
A recent analysis of Ref.~\cite{Wygas:2018otj} shows that pion condensation is unlikely to occur under this constraint if the lepton asymmetry is equally distributed among the three flavors.
However, due to the absence of neutrino oscillations at $T \gtrsim 10$~MeV, the individual lepton flavor asymmetries are not strongly constrained. It has been pointed out in~\cite{Wygas:2018otj,Middeldorf-Wygas:2020glx} that sufficient conditions for pion condensation to occur can be achieved for unequally distributed lepton asymmetries.
Complementary to~\cite{Middeldorf-Wygas:2020glx}, in the present letter we determine these conditions specifically~\cite{difference}. Moreover, we point out, for the first time, signatures of a pion-condensed phase in the early Universe,
namely its impacts on the spectrum of PGWs and on PBH formation.

\paragraph*{Equation of state.}

Pion condensation is expected to occur if the electric charge chemical potential $\mu_Q$ 
exceeds the pion mass.
First-principle lattice QCD studies at finite isospin density do suggest 
pion condensation to take place 
at $T \lesssim 160$~MeV and $\mu_I \gtrsim m_\pi$~\cite{Brandt:2017oyy,Brandt:2018bwq}, 
with 
$\mu_I$ being the isospin chemical potential~\cite{isovscharge}.
Here we analyze the cosmic trajectories determined by Eqs.~\eqref{eq:B}-\eqref{eq:l} at different values of $l_e$, $l_\mu$, and $l_\tau$ to determine the conditions for pion condensation to occur.
Notice that the weak decays of pions are blocked in the present 
setting of weak equilibrium, since all outgoing neutrino states are filled due to the high lepton chemical potentials, stabilizing the pion condensate~\cite{Abuki:2009hx}.

Neglecting QED interactions, the standard model equation of state is partitioned into contributions from QCD, leptons and photons:
\begin{align}
\label{eq:pSM}
p & =
p_{\rm QCD}(T,\mu_B,\mu_Q)
+
p_{L}(T,\mu_Q,\{\mu_\alpha\})
+
p_{\rm \gamma}(T)~.
\end{align}
The leptonic pressure is modeled by an ideal gas of charged leptons and neutrinos, including all three lepton flavors.
The photonic pressure is given by a massless ideal gas of photons.

\begin{figure}[t]
  \centering
  \includegraphics[width=.49\textwidth]{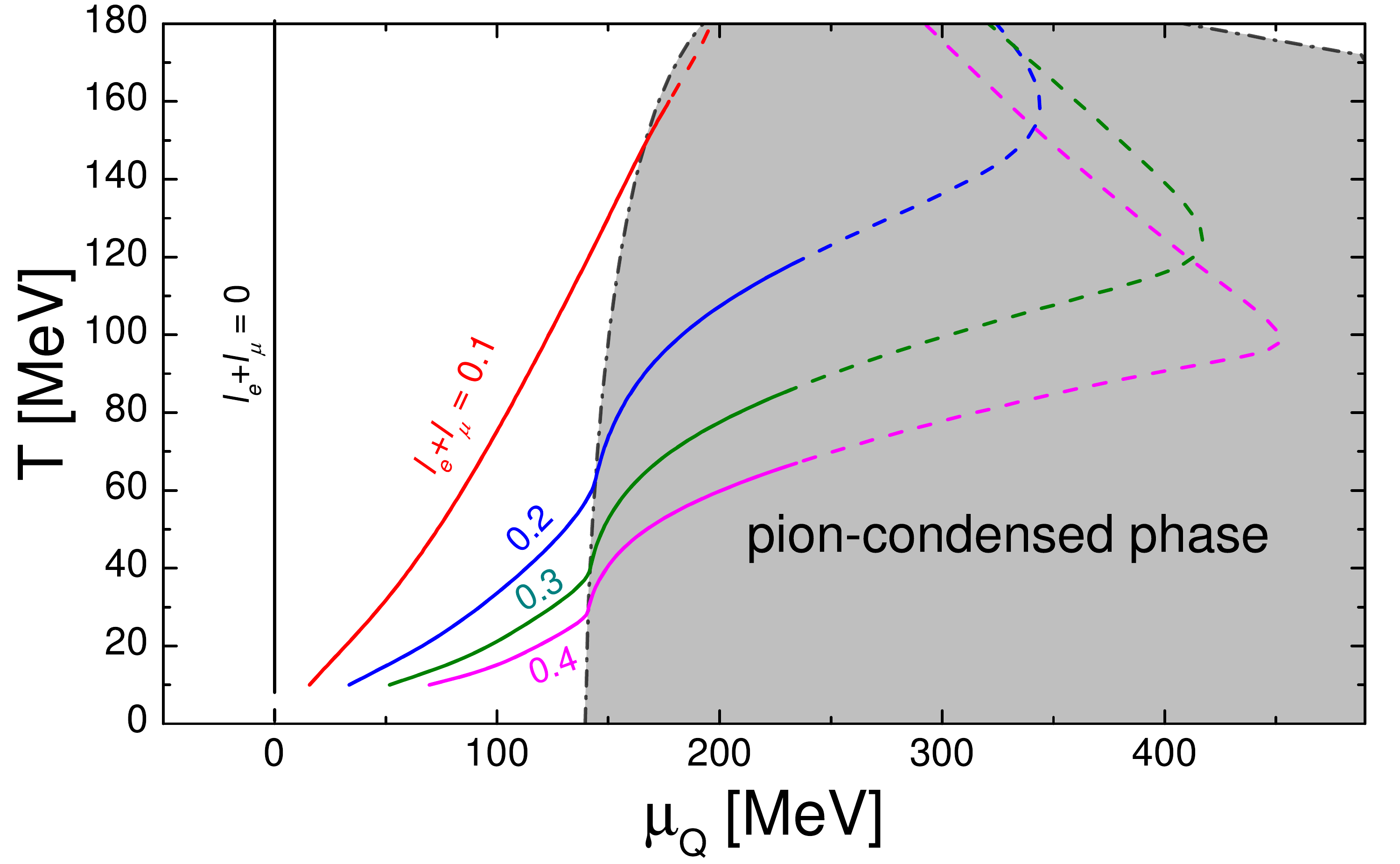}
  \caption{
  The phase diagram of an interacting hadron resonance gas with pion condensation in the $\mu_Q$-$T$ plane. 
  The dash-dotted line separates the pion condensed phase~(shaded area) from the normal phase.
  The colored lines depict cosmic trajectories for different values of the lepton flavor asymmetries: the standard cosmic trajectory~(black) and $l_e + l_\mu$ equal to 0.1~(red), 0.2~(blue), 0.3~(green), and 0.4~(magenta).
  In all cases $l_e = l_\mu$ and $l = l_e + l_\mu + l_\tau = 0$.
  The dashed parts of the trajectories correspond to regions where the effective mass model cannot be reliably validated with the lattice data.
  }
  \label{fig:pions}
\end{figure}

As we focus our study on temperatures $T < 160$~MeV that are relevant for hadronic matter,
the QCD pressure is approximated by a variant of the hadron resonance gas~(HRG) model. 
In the standard HRG model one includes all known  hadrons and resonances as free particles.
The HRG model  provides a reasonable description of the QCD equation of state in this temperature range when compared to the results of first-principle lattice QCD calculations~\cite{Borsanyi:2011sw,Bazavov:2012jq}.
To incorporate the pion-condensed phase we modify the HRG model by replacing the free pion gas by an interacting pion gas, modeled by a quasi-particle~(effective mass) approach~\cite{Savchuk:2020yxc} matched to chiral perturbation theory~\cite{Son:2000xc} and lattice QCD results at zero temperature~(see details in Ref.~\cite{EMM:sm}).
The reliability range of the model is established through comparisons to our first-principle lattice QCD results at $\mu_I>0$, as detailed in Ref.~\cite{lat:sm}.
The phase diagram of the model in $\mu_Q$-$T$ plane is shown in Fig.~\ref{fig:pions}.

The QCD pressure thus consists of the pressure of three pion species, each described by an effective mass model, and by contributions of the rest of the hadrons and resonances that are modeled as free particles:
\begin{equation}
\label{eq:pQCD}
p_{\rm QCD}(T,\mu_B,\mu_Q) = \sum_{i \in \pi} p_i^{\rm EM}(T,\mu_i) + \sum_{j} \, p_j^{\rm id}(T,\mu_j).
\end{equation}
Here $\mu_j = B_j \mu_B + Q_j \mu_Q$ with $B_j$ and $Q_j$ being the baryon and electric charge of hadron species $j$, respectively.
The index $i$ sums over the three pion species and the index $j$ sums over all hadrons excluding pions.
We include all established light flavored and strange hadrons listed in Particle Data Tables~\cite{Agashe:2014kda}.

All the conserved charge densities and the entropy density entering Eqs.~\eqref{eq:B}-\eqref{eq:l} are calculated as the corresponding derivatives of the pressure function~\eqref{eq:pSM}: $n_i = \partial p / \partial \mu_i$ for $i = B,Q,e,\mu,\tau$, and $s = \partial p / \partial T$.
For given values of the baryon and lepton asymmetries $b$ and $l_\alpha$, we evaluate the cosmic trajectory in the temperature range $10 < T < 180$~MeV by numerically solving Eqs.~\eqref{eq:B}-\eqref{eq:l} for the chemical potentials at each temperature.
The numerical solution is achieved using Broyden's method~\cite{Broyden}.
The procedure is implemented within an extended version of the open source \texttt{Thermal-FIST} package~\cite{Vovchenko:2019pjl}. We tested this procedure by reproducing the cosmic trajectories reported in Ref.~\cite{Wygas:2018otj} using the HRG model.

\paragraph*{Cosmic trajectories.}

We fix $b = 8.6 \cdot 10^{-11}$ and perform a parametric scan in $l_e$ and $l_{\mu}$.
As the restriction $|l| < 0.012$ on the total lepton asymmetry is rather strong we shall set $l_\tau = - (l_e + l_{\mu})$, meaning that we have a vanishing total lepton asymmetry~($l = 0$) in all our calculations.
For each value of $l_e$ and $l_\mu$, we start calculations at $T = 10$~MeV, where all cosmic trajectories are very similar, and gradually increase the temperature.
If the cosmic trajectory enters the phase with a Bose-Einstein condensate of pions, we register the temperature $T_{\rm cond}$ where the trajectory crosses the pion condensation boundary.

\begin{figure}[t]
  \centering
  \includegraphics[width=.49\textwidth]{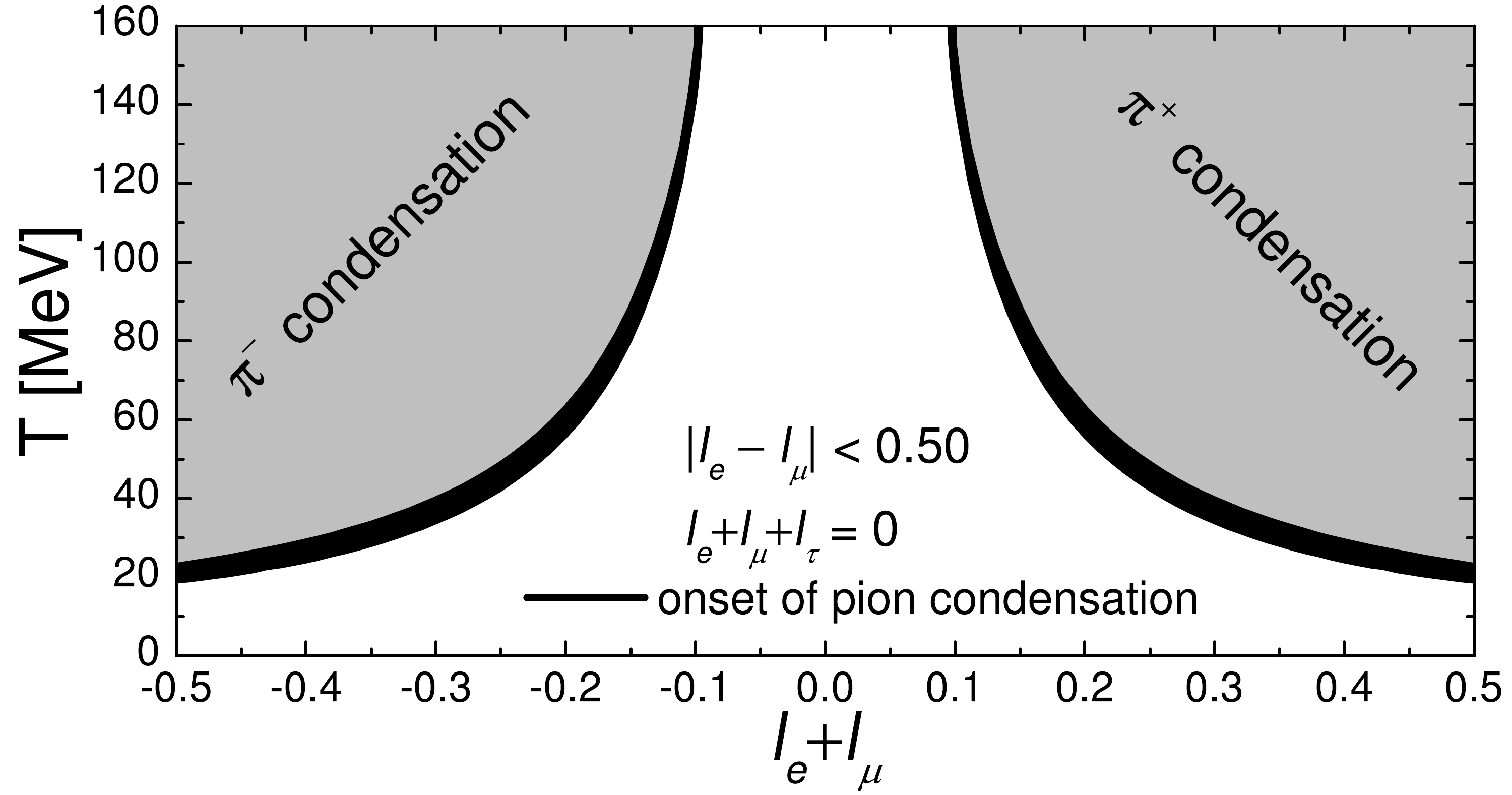}
  \caption{
  Dependence of the pion condensation onset temperature on the sum $l_e + l_{\mu}$ of electron and muon flavor asymmetries.
  The bands result from a variation of the difference of electron and muon asymmetries in a range $|l_e - l_{\mu}| < 0.50$.
  }
  \label{fig:scan}
\end{figure}

Our calculations reveal that $T_{\rm cond}$ depends mainly on the sum $l_e + l_\mu$ of the electron and muon lepton asymmetries, whereas the dependence on the difference $l_e - l_\mu$ is mild.
This is shown in Fig.~\ref{fig:scan}, where we depict the dependence of the temperature $T_{\rm cond}$ on
the sum $l_e + l_\mu$. The difference $l_e - l_\mu$ is varied in a range $|l_e - l_\mu| < 0.5$, giving the narrow black uncertainty band in Fig.~\ref{fig:scan}.
At temperatures between $T_{\rm cond}$ and the chiral crossover pseudocritical temperature $T_{\rm pc} \approx 160$~MeV the cosmic matter is in a pion-condensed phase.
We find that pion condensation occurs in the early Universe at $T < 160$~MeV if the following condition is met:
\begin{equation}
\label{eq:crit}
|l_e + l_{\mu}| \gtrsim 0.1~.
\end{equation}
Pion condensation is not observed at smaller absolute values of $l_e + l_{\mu}$.
The relation \eqref{eq:crit} can therefore be regarded as a universal criterion for pion condensation in the early Universe.
Positive values of $l_e + l_{\mu}$ correspond to $\pi^+$ condensation, while negative $l_e + l_{\mu}$ imply $\pi^-$ condensation.

\begin{figure}[t]
  \centering
  \includegraphics[width=.49\textwidth]{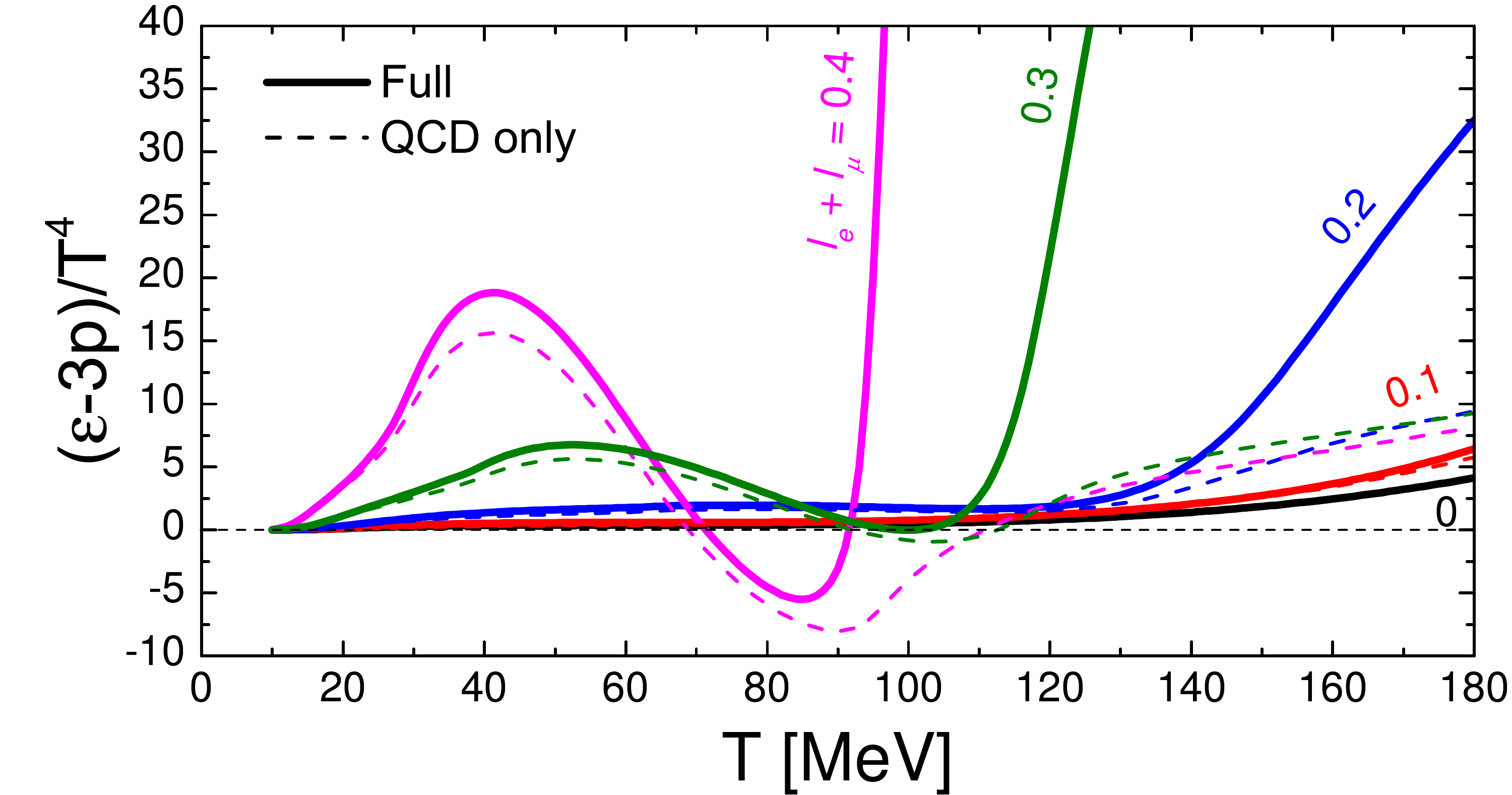}
  \caption{
  Temperature dependence of the interaction measure, $(\varepsilon-3p)/T^4$ along the cosmic trajectory for different values of the lepton flavor asymmetries: $l_e + l_\mu = 0$~(black), 0.1~(red), 0.2~(blue), 0.3~(green), and 0.4~(magenta). In all cases $l_e + l_\mu + l_\tau = 0$.
  }
  \label{fig:IT4}
\end{figure}

The temperature dependence of $\mu_Q$ is shown in Fig.~\ref{fig:pions} for several different values of lepton flavor asymmetries in the range $0\le l_e + l_\mu \le 0.4$. These values are motivated by various theoretical predictions to explain the baryon and lepton asymmetry in the early universe, see Refs.~\cite{Affleck:1984fy,Stuke:2011wz,Casas:1997gx,McDonald:1999in,Abazajian:2004aj,Ichikawa:2004pb}.
For $l_e + l_\mu = 0$ one essentially recovers the standard cosmological trajectory where $\mu_Q$ is very close to zero throughout and far away from the pion condensed phase.
For sufficiently large absolute values of $l_e + l_\mu$~[see Eq.~\eqref{eq:crit}], the cosmic trajectory crosses the pion condensation boundary.
The kink-like structure in the cosmic trajectory, predominantly visible for the $l_e + l_\mu = 0.4$ case at $T \approx 95$~MeV, is associated with a rapid growth of the lepton chemical potentials.

The equation of state exhibits an interesting behavior for trajectories that enter the pion condensed phase.
Of particular interest is the interaction measure, $(\varepsilon - 3p)/T^4$.
The interaction measure is negative deep in the pion-condensed phase at moderate temperatures~(see Fig.~\ref{fig:IT4}) -- a distinctive feature of the pion condensed phase also seen in lattice QCD calculations.
Figure~\ref{fig:IT4} depicts the temperature dependence of $(\varepsilon - 3p)/T^4$ 
along the cosmic trajectory for the four different cases of positive $l_e + l_\mu$ values discussed above.
The behavior of these two quantities is significantly affected at large lepton asymmetries.
For $|l_e + l_\mu| \gtrsim 0.3$ the cosmic trajectory passes through a region with negative $(\varepsilon - 3p)/T^4$, as illustrated by the magenta curve in Fig.~\ref{fig:IT4} for $l_e + l_\mu = 0.4$. 
Negative interaction measure correlates with large sound velocities that go above the conformal limit of $c_s^2 = 1/3$.
The interaction measure grows to large values $(\varepsilon - 3p)/T^4 \gtrsim 10$ at larger temperatures.
This drastic rise is a consequence of large lepton chemical potentials at these temperatures, which emerge from lepton flavor number conservation.

\paragraph*{Effects on the spectrum of PGWs.}
 Due to  the presence of a nonvanishing lepton asymmetry and the possible formation of the pion-condensed phase,  the equation of state before big bang nucleosynthesis~(BBN) can change, which will leave an imprint on the PGW spectrum~\cite{Hajkarim:2019csy,Hajkarim:2019nbx,Saikawa:2018rcs,Schettler:2010dp}.

The evolution of each polarization $\lambda$ of tensor perturbation $h$ for a mode $k$ in cosmology is given by \cite{Mukhanov:1990me,Mukhanov:2005sc}
\begin{eqnarray}
\label{tenpert}
h_{{\bf k},\lambda}^{\prime\prime}+2\frac{a^\prime}{a}h_{{\bf k},\lambda}^\prime+k^2h_{{\bf k},\lambda}=0\,,
\end{eqnarray}
where the $^\prime \equiv d/d\eta$ is the derivative with respect to conformal time $\eta$ and $a$ is the scale factor ($ad\eta=dt$, $t$ is the cosmic time).  The primordial tensor perturbation can be written in terms of the transfer function $X$, tensor perturbation amplitude $h_{\bf k,\lambda}^{prim}$ and  tensor power spectrum parameterized 
with respect to a characteristic scale $\tilde{k}=0.05$~Mpc$^{-1}$  
 \begin{eqnarray}
 \label{xhper}
 h_{\bf k,\lambda}(\eta)\equiv  h_{\bf k,\lambda}^{prim}X(k,\eta),\,\, \mathcal{P}_T=\sum_\lambda |h^{prim}_{\bf k,\lambda}|^2=A_T\left(\frac{k}{\tilde{k}}\right)^{n_T},
 \end{eqnarray}
 where $A_T=r\,A_S$ and $A_S$, $n_T$ are scalar and tensor perturbation amplitudes, and the tensor spectral index, respectively. The tensor to scalar ratio denoted by  $r$ has an upper limit from measurements by PLANCK of $r\lesssim 0.07$ \cite{Akrami:2018odb,Aghanim:2018eyx}.

To compute the temporal evolution of the scale factor one needs to solve the Friedmann equation ($H^2=({\dot a}/a)^2=(8\pi/3M_{Pl}^2)\varepsilon$, $M_{Pl}=1.22\times10^{19}$~GeV). 
We solve Eq.~(\ref{tenpert}) for a mode $k$ using (\ref{xhper}) until horizon 
 crossing~\cite{firstfootnote},
 i.e. when $k=|{\bf k}|=a(\eta_h) H(\eta_h)$, then we  use the WKB (Wentzel, Kramers, Brillouin) approximation for the PGW afterwards until today \cite{Watanabe:2006qe,Bernal:2019lpc}.
Using Eqs.~(\ref{tenpert}) and (\ref{xhper}) the relic density of PGWs for different frequencies $\nu=k/2\pi$ at today ($a_0$) can be computed from \cite{Watanabe:2006qe,Bernal:2019lpc}
\begin{eqnarray}
\label{omrelic}
\Omega_{\text{GW}}(k,\eta_0)=\frac{\mathcal{P}_T(k) \left[X^\prime(k,\eta_0)\right]^2}{24a_0^2H_0^2}\,.
\end{eqnarray}

\begin{figure}[t]
  \centering
  \includegraphics[width=.49\textwidth]{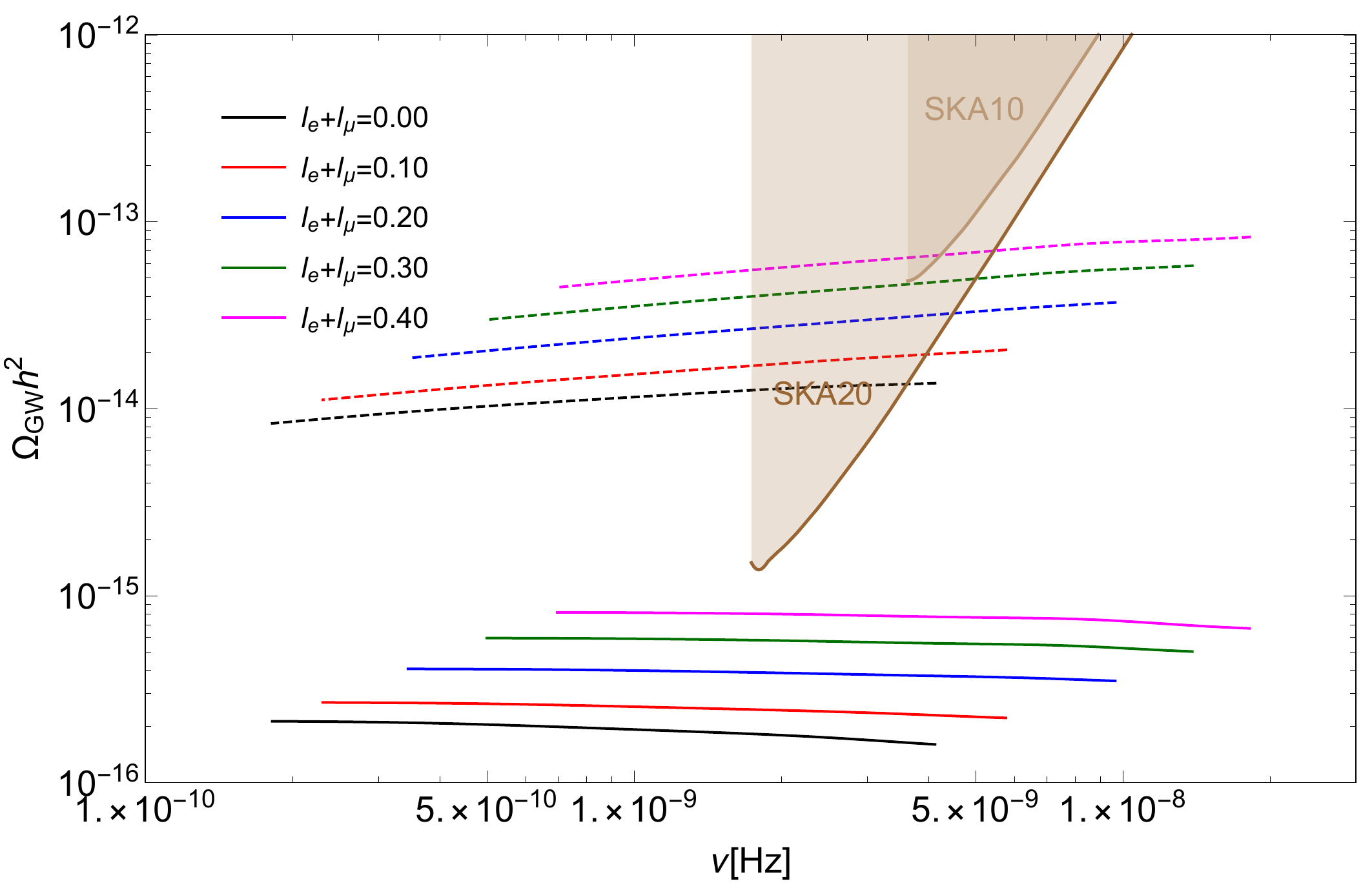}
  \caption{
  PGW relic density for different lepton asymmetry values and using the amplitude of scalar perturbation  $A_S=2.1\times 10^{-9}$, the scale invariant $n_T=0$ (solid lines) and the scale dependent $n_T=0.25$ (dashed lines) tensor power spectrum from the upper bound on the tensor to scalar perturbation ratio $r=0.07$ of PLANCK. The future constraints that can be reached by the SKA over 10 and 20 years of operation are also shown~\cite{Janssen:2014dka,Bull:2018lat}.
  }
  \label{fig:pgw}
\end{figure}

Using the  
equations of state computed for different lepton asymmetry values, for which the cosmic trajectory can enter the pion condensed regime, one can estimate the PGW spectrum by using Eqs.~(\ref{tenpert})-(\ref{omrelic}). 
We consider entropy conservation ($s\,a^3=const.$) 
and use the number of degrees of freedom after neutrino decoupling  \cite{Drees:2015exa} to find the relation between the scale factor and the temperature.  The PGW relic spectra are shown in Fig.~\ref{fig:pgw}. As the lepton asymmetry increases, so does the amplitude of the spectrum because the entropy, energy and pressure densities become larger. 
Moreover, the formation of pion condensation can enhance the PGW due to the change of equation of state.
Pulsar timing arrays, such as the Square Kilometre Array~(SKA)~\cite{Janssen:2014dka,Bull:2018lat}, can measure the predicted PGW spectrum especially around the QCD phase transition if it is scale invariant ($n_T=0$) or blue-tilted ($n_T>0$). The LISA experiment \cite{Audley:2017drz} can also measure such effects at higher frequencies. The lepton asymmetry at BBN time and afterwards is constrained by cosmic microwave background measurements. Since nonvanishing lepton asymmetry and pion condensation before BBN can modify the PGW spectrum, GW observatories with high sensitivity  are able to measure these effects in the early Universe.

\paragraph*{Impact on the formation of PBHs.}

\begin{figure}[t]
  \centering
  \includegraphics[width=.49\textwidth]{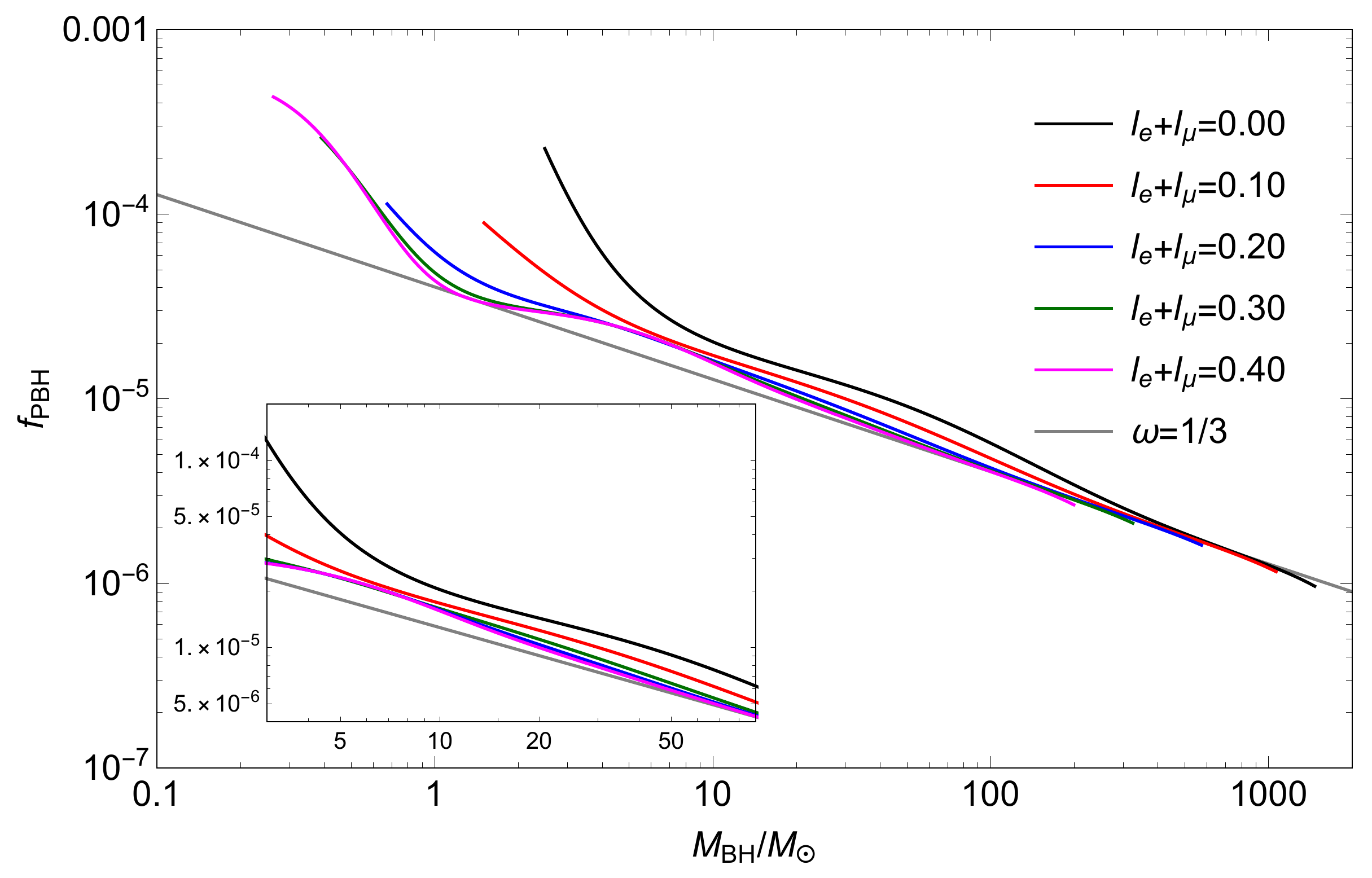}
  \caption{
The fraction of PBHs with respect to PBH masses for different lepton asymmetry values (different colors) assuming the scale invariant Gaussian  density perturbation and the value of the density spectral index $n_M= 0$~\cite{pbhform:sm}. The gray line denotes the result for a purely radiation dominated background fluid. 
  } \label{fig:pbh}
\end{figure}

The population of primordial black holes that formed in the early Universe depends on the Hubble rate and the  total mass within the Hubble horizon  \cite{Byrnes:2018clq,Widerin:1998my,Sobrinho:2016fay,Jedamzik:1996mr,Schmid:1998mx,Niemeyer:1997mt,Niemeyer:1999ak}. 
As mentioned earlier, a nonvanishing lepton asymmetry and a pion condensed phase modify the Hubble rate thereby modifying the production of PBHs in specific range of masses. 
The horizon mass, defined as
$M_{h}=\frac{4\pi}{3}H^{-3}\varepsilon\,$~ \cite{Carr:1975qj,Carr:1974nx},
relates a given temperature in the early Universe to the horizon mass and later on to a typical black hole mass  $M_{\text{BH}}$. 
Figure~\ref{fig:pbh} shows the fraction $f_{\text{PBH}}$ of PBHs with respect to total cold dark matter (CDM) abundance for different lepton asymmetry cases~(see Ref.~\cite{pbhform:sm} for the technical details of the calculation).
The presence of pion condensation is signalled by a modification of $f_{\text{PBH}}$ at masses larger than one solar mass.

The parameter $f_{\text{PBH}}$ can be indirectly  measured by different experiments. The fraction of PBHs with masses  $10^{-6}  M_{\odot} \lesssim M_{\text{BH}}\lesssim 10^3 M_{\odot}$ from some experimental constraints  (OGLE, HSC, Caustic, EROS, MACHO) should be  $f_{\text{PBH}} \lesssim 0.05$ 
\cite{secondfootnote,Niikura:2017zjd,Niikura:2019kqi,Tisserand:2006zx,Allsman:2000kg,Oguri:2017ock}.  
The SKA \cite{Janssen:2014dka,Bull:2018lat} and LISA \cite{Audley:2017drz} can also indirectly constrain the fraction of PBHs by putting limits on the induced PGWs from curvature perturbation or  using GWs produced by coalescing events \cite{Wang:2019kaf,Bartolo:2018rku,Hajkarim:2019csy}.

\vspace{0in}

\paragraph*{Summary.}

The present analysis of cosmic trajectories at non-vanishing lepton flavor asymmetries reveals a simple criterion for the onset of pion condensation in the early Universe -- it occurs when the total electron and muon asymmetry parameter is sufficiently large, $|l_e + l_\mu| \gtrsim 0.1$.
This result does not exhibit large sensitivity to the modeling of pion interactions.
Asymmetries beyond this value lead the system deep inside the pion condensed phase,
affecting its equation of state considerably.
The possible presence of such a Bose-Einstein condensed phase of pions would have significant cosmological implications such as the strong enhancement of the spectrum of PGWs and the change of the fraction of PBHs with mass larger than one solar mass. The experimental signatures of pion condensation from the early Universe can be probed by pulsar timing and GW detectors. The recent  BHs merger event of LIGO GW190521 can be from PBHs produced during the pion condensation epoch \cite{Abbott:2020mjq,Abbott:2020tfl}.

Pion condensation could also affect big bang nucleosynthesis.
If the pion condensed phase is present, spheres of pions and leptons -- the pion stars -- can form which are stabilized by the high density of neutrinos due to the high lepton chemical potentials~\cite{Carignano:2016lxe,Brandt:2018bwq,Andersen:2018nzq}.
Typical pion star masses will be in the range of a few solar masses when the early Universe leaves the pion condensed phase. 
The neutrinos will diffuse out of the pion stars on the timescale of weak interactions. The situation is similar to the one for proto-neutron stars where neutrinos leave on the timescale of several seconds. Hence, pion stars would decay around the time of BBN. 
The produced high energy leptons would influence the abundances of primordially produced nuclei, which could be addressed by a modified BBN simulation.


\begin{acknowledgments}

\emph{Acknowledgments.} 
We thank Szabolcs Bors\'anyi for useful correspondence and for providing the data for the Taylor expansion coefficients.
We also thank Dietrich B\"odeker, Eduardo Fraga, Mauricio Hippert, Pasi Huovinen, Mandy M. Middeldorf-Wygas, Isabel Oldengott, Sebastian Schmalzbauer, Dominik Schwarz, Stephan Wystub, and Yong Xu
for numerous fruitful discussions. 
V.V. was supported by the
Feodor Lynen program of the Alexander von Humboldt
foundation and by the U.S. Department of Energy, 
Office of Science, Office of Nuclear Physics, under contract number  DE-AC02-05CH11231231. 
The work of B.B.B., F.C., G.E., F.H. and  J.S. is supported by the Deutsche Forschungsgemeinschaft (DFG) through the CRC-TR 211, project number 315477589-TRR 211.
G.E.\ also acknowledges support by the DFG Emmy Noether
Programme (EN 1064/2-1).
The work of F.H. is also supported by the research grant “New Theoretical Tools for Axion Cosmology” under the Supporting TAlent in ReSearch@University of Padova (STARS@UNIPD).

\end{acknowledgments}

\bibliography{PionCondensation}

\begin{thebibliography}{94}
\bibitem [{fir()}]{firstfootnote}%
  \BibitemOpen
  \href@noop {} {}\bibinfo {note} {The initial conditions that we consider at
  superhorizon scale ($k\ll aH$) are $X(k,0)=1$ and
  $X^\prime(k,0)=0$.}\BibitemShut {Stop}%
\end{thebibliography}

\section*{Supplemental material}

\setcounter{secnumdepth}{2}

\section{Effective mass model for pion condensation}
\label{app:EMM}

We use a quasiparticle~(effective mass) approach to describe interacting pions with a pion-condensed phase.
Outside of the pion condensed phase, the pressure of a single pion species in the effective mass model reads~\cite{Savchuk:2020yxc}
\eq{\label{eq:pEM}
p_\pi^{\rm EM}(T,\mu_\pi; m^*) = p^{\rm id}_\pi(T,\mu_\pi;m^*) + p_f(m^*).
}
Here $\pi \in \pi^+, \pi^-, \pi^0$.
The rearrangement term $p_f(m^*)$ is a consequence of interactions.
It ensures a proper counting of the interaction energy
and preserves the thermodynamic consistency in the quasiparticle model. 
For instance, it ensures that the quasiparticle pion number density, $n_\pi^{\rm EM} = n^{\rm id}_\pi(T,\mu_\pi;m^*)$, satisfies a thermodynamic relation $n_\pi^{\rm EM} = (\partial p_\pi^{\rm EM} / \partial \mu_\pi)_{T}$, correctly taking into account the medium dependence of the effective mass, $m^*(T,\mu)$.
The specific form of $p_f(m^*)$ defines the quasiparticle model.
Here we take $p_f(m^*)$ in the form
\eq{\label{eq:chPTpf}
p_f(m^*) = \frac{(m^*)^2 f_\pi^2}{4} \, \left[ 1 - \frac{m_\pi^2}{(m^*)^2} \right]^2~,
}
chosen to match the model to chiral perturbation theory and lattice QCD results in the pion-condensed phase at $T = 0$~(see below).
The pressure at a given $T$ and $\mu_\pi$ has to be maximized with respect to $m^*$, resulting in a gap equation $(\partial p_\pi^{\rm EM} / \partial m^*)_{T,\mu} = 0$:
\eq{\label{eq:gap}
p_f'(m^*) = n^{\rm id}_\sigma(T,\mu_\pi;m^*)~.
}
Here $n^{\rm id}_\sigma(T,\mu_\pi;m^*) \equiv -\partial p_\pi^{\rm id}/\partial m^*$ is the scalar density of an ideal gas of pions with mass $m^*$.
A numerical solution to the gap equation determines $m^*$ at given $T$ and $\mu_\pi$, allowing to calculate all other thermodynamic quantities through Eq.~\eqref{eq:chPTpf}.

\begin{figure}[h]
  \centering
  \includegraphics[width=.48\textwidth]{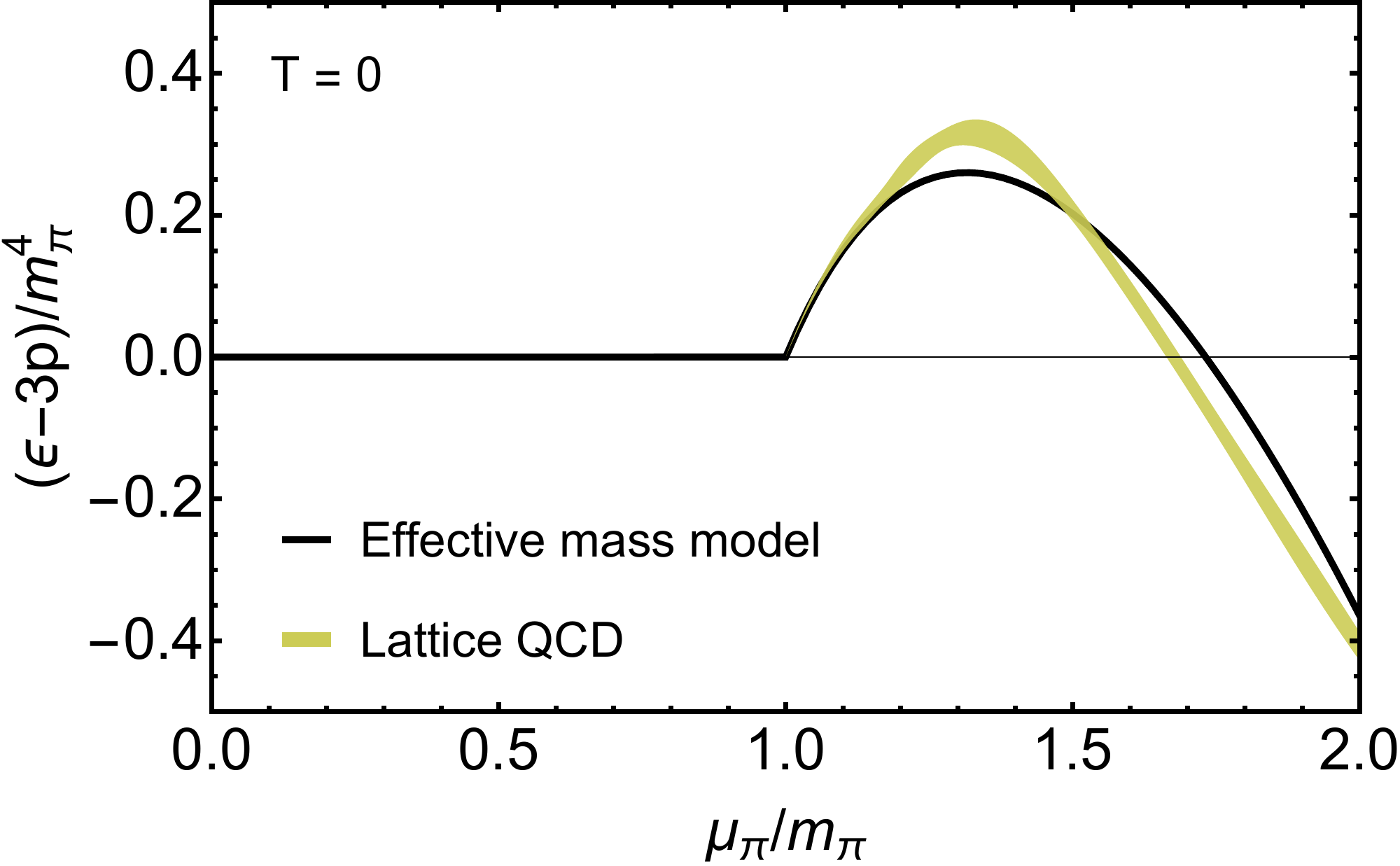}
  \caption{
  The dependence of the normalized trace anomaly $(\varepsilon-3p)/m_\pi^4$ on the normalized pion chemical potential $\mu_\pi/m_\pi$, evaluated in the effective mass model at $T = 0$.
  The yellow band depicts lattice QCD results from Ref.~\cite{Brandt:2018bwq}.
  }
  \label{fig:pionstrace}
\end{figure}

The transition to the pion-condensed phase takes place when the effective pion mass becomes equal to the chemical potential, $m^* = \mu_\pi$.
The equation determining the transition line in the $\mu_\pi$-$T$ plane reads~\cite{Savchuk:2020yxc}
\eq{\label{eq:muc}
p_f'(\mu_\pi) = n^{\rm id}_\sigma(T,\mu_\pi;m^* = \mu_\pi)~.
}

The effective mass equals the chemical potential in the phase diagram region with a pion condensate, $m^* = \mu_\pi$ for $\mu_\pi \geq \mu_{\rm cond}$, as a consequence of interactions between thermal and condensed pions~\cite{Barz:1989cv}. Here $\mu_{\rm cond}$ is the pion chemical potential at pion condensation boundary.
Therefore, the pressure in this phase reads reads
\eq{\label{eq:pBEC}
p_\pi^{\rm EM}(T,\mu_\pi) = p^{\rm id}(T,\mu_\pi;m^* = \mu_\pi) + p_f(\mu_\pi).
}
At $T = 0$, the pion number density $n_\pi^{\rm EM} = (\partial p_\pi^{\rm EM} / \partial \mu_\pi)_T$ reads
\eq{\label{eq:npiChPT}
n_\pi^{\rm EM}(T=0,\mu_\pi) 
& = p_f'(\mu) \, \theta(\mu - m_\pi)
\nonumber \\
& = \frac{\mu_\pi \, f_\pi^2}{2} \left[1 - \frac{m_\pi^4}{\mu_\pi^4} \right] \, \theta(\mu - m_\pi).
}
Equation~\eqref{eq:npiChPT} matches the result of leading-order chiral perturbation theory~\cite{Son:2000xc}, which for $f_\pi = 133$~MeV describes well the available lattice QCD data on isospin density at $T = 0$~\cite{Brandt:2018bwq}. Recently, these chiral perturbation 
theory predictions have been backed up by next-to-leading-order 
calculations, both for the density and for the equation of state~\cite{Adhikari:2019zaj,Adhikari:2019mlf,Adhikari:2020ufo,Adhikari:2020kdn}.

\section{Lattice simulations}
\label{app:latsim}

Here we describe the details of our 
first-principles lattice QCD simulations at nonzero 
isospin density. On the one hand, the lattice results at (approximately) 
zero temperature are used to guide the construction of the effective mass model
described above.
Here we use our
data at a single lattice spacing from Ref.~\cite{Brandt:2018bwq}. 
On the other hand, the finite-temperature results serve to test the validity range of 
the model at nonzero isospin and zero baryon density.
To this end we employ our data from Refs.~\cite{Brandt:2017oyy,Brandt:2018omg} 
on four lattice spacings.

To simulate the path integral $\mathcal Z$
we take the tree-level Symanzik-improved 
gauge action and $2+1$ flavors of rooted staggered quarks 
with physical masses~\cite{Borsanyi:2010cj}. 
The isospin chemical potential $\mu_I$ enters the Dirac 
operator\footnote{This convention, for which pion condensation sets in at $\mu_I=m_\pi$ at 
zero temperature, differs from that used in our earlier works~\cite{Brandt:2017oyy,Brandt:2018omg} by a factor of two.}
via the quark chemical potentials $\mu_u=-\mu_d=\mu_I/2$, while 
$\mu_s=0$. Comparing to the standard basis with baryon and charge chemical 
potentials, one can read off $\mu_Q=\mu_I$, $\mu_B=-\mu_I/2$. The 
simulations therefore correspond to a situation with a specific linear 
combination of baryon and charge chemical potentials, which only couples 
to hadron species containing an unequal number of up and down quarks (predominantly charged pions).\footnote{Note that
the baryon density still vanishes in our simulations: it is obtained in terms of derivatives with respect to the quark chemical potentials as $n_B=n_u/3+n_d/3+n_s/3=0$ at pure isospin chemical potential, where $n_u=-n_d$ and $n_s=0$.} 
To be able to perform the simulations, we further need to introduce an auxiliary pionic source $\lambda>0$ that is extrapolated to zero at the end of the analysis.
The role of the $\lambda$ parameter is twofold. First, it triggers the spontaneous symmetry breaking corresponding to pion condensation in a finite volume. Second, it serves to stabilize 
the theory in the infrared by making the Goldstone boson of
the pion condensed phase slightly 
massive~\cite{Brandt:2017oyy}.

To calculate the equation of state, 
our primary observable is the isospin density
\begin{equation}
n_I(T,\mu_I) = \frac{T}{V}\frac{\partial \log\mathcal Z}{\partial \mu_I}\,.
\end{equation}
The details of the $\lambda\to0$ extrapolation of this observable are explained in Ref.~\cite{Brandt:2018omg} and in the following 
we work with the so extrapolated quantity.
From $n_I$, we can calculate $\Delta \mathcal O(T,\mu_I)\equiv \mathcal O(T,\mu_I)-\mathcal O(T,0)$ for any observable $\mathcal O$.
In particular, the pressure difference and the 
trace anomaly difference
can be constructed as
\begin{align}
 \Delta p(T,\mu_I) &= \int_0^{\mu_I}\!\!\!\! \textmd{d}\mu_I' \,n_I(T,\mu_I')\,, \\
\Delta I(T,\mu_I) &= \mu_I n_I(T,\mu_I)\!+\!\int_0^{\mu_I} \!\!\!\! \textmd{d}\mu_I' \left(T\frac{\partial}{\partial T} -4\right) n_I(T,\mu_I') \,. \label{eq23}
\end{align}

The zero-temperature results for $n_I$ near $\mu_I=m_\pi$ 
are well-described by the chiral perturbation theory 
formula~\eqref{eq:npiChPT} with $f_\pi=133(4)\textmd{ MeV}$~\cite{Brandt:2018bwq,Adhikari:2019zaj,Adhikari:2019mlf,Adhikari:2020ufo,Adhikari:2020kdn}. This is smoothly matched by 
a spline interpolation for $n_I(\mu_I)$ at higher values of 
the chemical potential. The interaction measure is determined 
via Eq.~\eqref{eq23} -- note that at zero temperature $\Delta I = I$ and, moreover, the first contribution to the integral in $\Delta I$ of Eq.~\eqref{eq23} vanishes, simplifying 
this expression considerably. The so obtained curve is plotted in 
Fig.~\ref{fig:pionstrace} as the yellow band.

\begin{figure}[b]
  \centering
  \includegraphics[width=.48\textwidth]{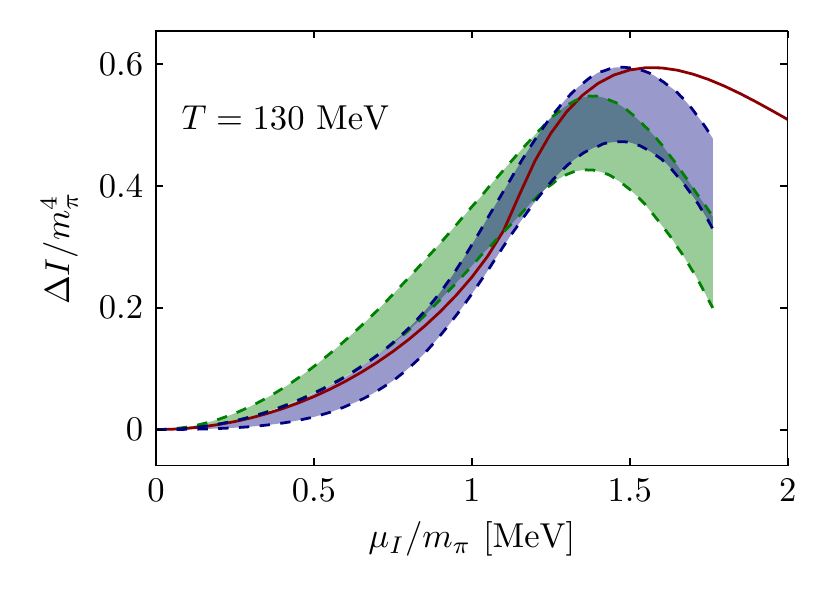}
  \includegraphics[width=.48\textwidth]{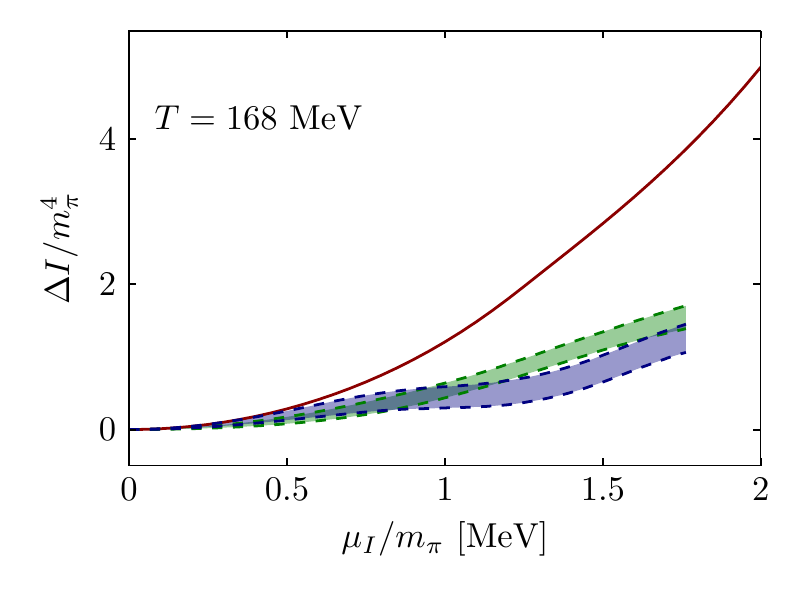}
  \caption{
  The trace anomaly difference as a function of $\mu_I$ at two 
  different temperatures on our $N_t=10$ (green) and $N_t=12$ 
  (blue) lattice ensembles, compared to the effective mass model
  (red curve). 
  }
  \label{fig:deltaI_plot}
\end{figure}

For testing the effective mass model at $T>0$ we 
concentrate on $\Delta I$ because compared to 
other observables it is found to contain
the least amount of lattice discretization errors\footnote{Note that this choice allows to discuss the $\mu_I$-dependence of the model 
but not its reliablility at $\mu_I=0$. However, for the effect of 
pion condensation on the cosmic trajectory, we expect the latter 
to be less important.}. 
The integrals and the derivatives in Eq.~\eqref{eq23} need to be evaluated numerically. To this end we fit 
$n_I(T,\mu_I)$ via a two-dimensional spline surface. The spline nodepoints 
are drawn from a Monte-Carlo procedure with the goodness 
of the fit playing the role of the action, providing 
a direct estimate of systematic errors (see Ref.~\cite{Brandt:2016zdy} for more details). The $\mu_I$-dependence of 
$\Delta I$ is plotted 
for two representative values of the temperature in Fig.~\ref{fig:deltaI_plot}. Here we include the results for 
our two finest lattice spacings, $N_t=10$ and $N_t=12$. (The continuum
limit at constant $T$ corresponds to $N_t\to\infty$, but we do 
not carry out this extrapolation here.)
The model is found to capture the notable features
of the lattice data qualitatively. A quantitative description is obtained if 
neither $T$ nor $\mu_I$ are too large. In particular, sizeable deviations are 
visible above the chiral restoration temperature, because the effective mass 
model does not contain the details of the physics of this phase transition.

\begin{figure}[h]
  \centering
  \includegraphics[width=.48\textwidth]{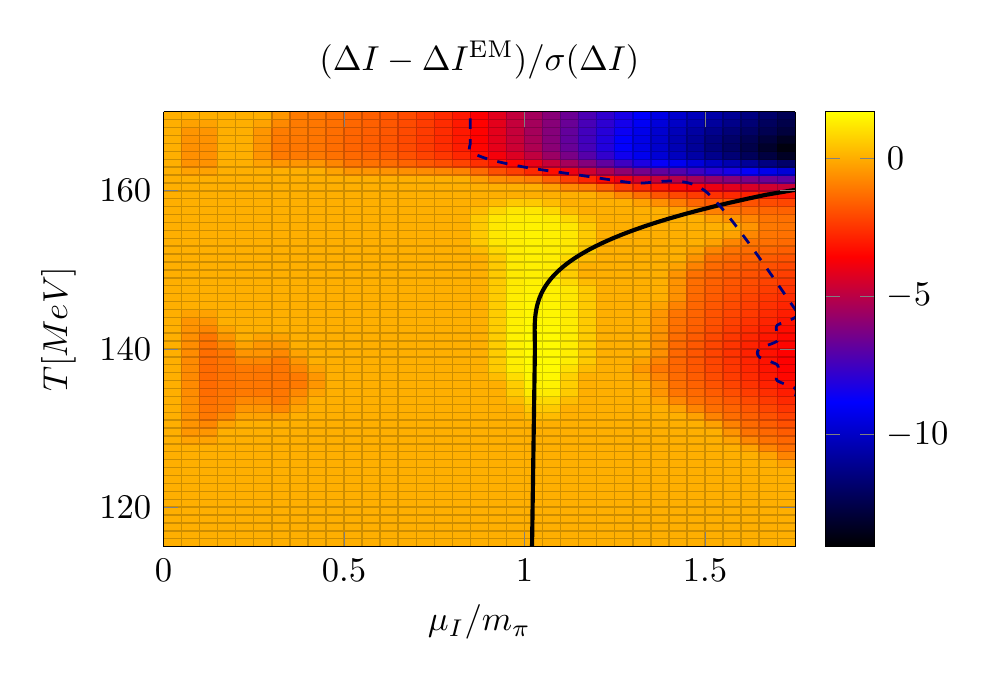}
  \caption{
  Heat plot of the deviation between the effective mass
  model and the lattice results for the trace anomaly difference $\Delta I$. For the latter our finest, $N_t=12$ ensembles 
  are used. The deviation is normalized by the 
  error of the lattice data. The solid black line indicates 
  the lattice result for the pion condensation boundary, while 
  the dashed line denotes the contour of $3$.
  }
  \label{fig:deltaI_surf}
\end{figure}

To make the comparison between the $N_t=12$ lattice results and the 
model more systematic, in Fig.~\ref{fig:deltaI_surf}
we show the deviation between the two in the form of a heat plot. 
Here we normalize by the error $\sigma$ of the lattice results -- therefore a
value of $n$ indicates a difference by $n$ standard deviations. 
The plot shows substantial differences for $\mu_I>m_\pi$ at high temperatures as well as slight deviations near the boundary of 
the pion condensed phase. We take the contour line at $3$ 
standard deviations as a marker and consider the model 
reliable in the parameter range where
\begin{equation}
 \frac{|\Delta I - \Delta I^{\rm EM}|}{\sigma (\Delta I)} \le 3\,,
\end{equation}
with $\Delta I^{\rm EM}$ being the subtracted interaction measure in the effective mass model.
This range is indicated by the solid line sections of the cosmic trajectories
in Fig.~1 of the main text.

The above comparisons were performed at nonzero isospin chemical 
potential $\mu_I$, where lattice results are available. 
For the analysis of the cosmic trajectory, the 
model is employed instead at nonzero charge chemical potential 
$\mu_Q$ (as well as low baryon chemical potential $\mu_B$).
At zero temperature, $\mu_I$ and $\mu_Q$ can be identified as long 
as the only charged states that contribute to the equation of state
have zero strangeness and zero baryon number. This is the case 
for $\mu_I<m_K$ (even in this case, kaon 
condensation is not expected to occur if a pion condensate is 
already present~\cite{Mannarelli:2019hgn}) and sufficiently 
low $\mu_B$ as is the case for the parameters considered 
in this paper.

\begin{figure}[t]
  \centering
  \includegraphics[width=.48\textwidth]{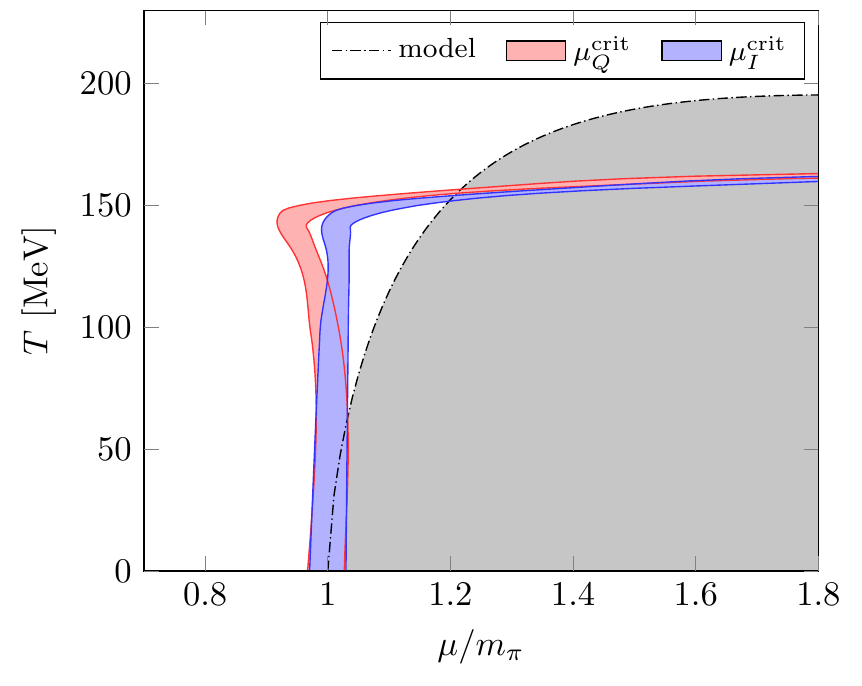}
  \caption{
  The critical lines $\mu_Q^{\rm crit}$ and $\mu_I^{\rm crit}$ as obtained on $N_t=12$ lattices from the approximation in Eq.~\eqref{clApprox} and from direct simulations, respectively. The critical line from the effective mass model is included for comparison.
  }
  \label{fig:boundariesNt12}
\end{figure}

Contrary to the identification $\mu_I=\mu_Q$ at zero temperature, 
for $T>0$ the different couplings of the two chemical potentials 
to hadronic states becomes relevant and the equation of state 
differs in the two cases. Nevertheless, in the effective
mass model the pion condensation boundary expressed in $\mu_I$ or 
in $\mu_Q$ remains the same, because interactions between pions and 
other hadrons are neglected in the model. 
The difference between the critical lines, $\mu_Q^{\rm crit}(T)$ 
and $\mu_I^{\rm crit}(T)$ can be estimated using 
lattice results for the estimators of the convergence radii of the corresponding Taylor series around $\mu_Q=\mu_I=0$.
In particular, we consider the expansions of the pressure,
\begin{equation}
\frac{p}{T^4} = \frac{c_2^{I,Q}}{2} \left(\frac{\mu_{I,Q}}{T}\right)^2 +\frac{c_4^{I,Q}}{24} \left(\frac{\mu_{I,Q}}{T}\right)^4 + \ldots
\end{equation}
and the estimators for the convergence radius for the 
susceptibilities $\chi_{I,Q} =\partial^2 p/\partial \mu_{I,Q}^2$.
We use the Taylor coefficients determined in Ref.~\cite{Borsanyi:2011sw} for our action and lattice spacings.
The leading estimator 
\begin{equation}
\frac{r_2(\chi_{I,Q})}{T}  = \sqrt{\frac{2c_2^{I,Q}}{c_4^{I,Q}}}\,,
\end{equation}
for the isospin direction was found to give a remarkably 
good approximation to the true critical line, $\mu_I^{\rm crit}$~\cite{Brandt:2018omg}. 
We assume this is also the case for the expansion in $\mu_Q$. Thus 
we approximate the critical line in the electric charge direction by
\begin{equation}
\mu_Q^{\rm crit} \approx \mu_I^{\rm crit} \cdot \frac{r_2(\chi_Q)}{r_2(\chi_I)}\,. \label{clApprox}
\end{equation}
For the second factor we use the lattice results~\cite{Borsanyi:2011sw} above and ideal HRG with quantum statistics below a matching temperature of $105 \textmd{ MeV}$.
The $N_t=12$ results for this approximation, together with the corresponding directly determined isospin 
critical line $\mu_I^{\rm crit}$~\cite{Brandt:2017oyy,Brandt:2018omg}, are plotted in Fig.~\ref{fig:boundariesNt12}.

Comparing to the effective mass model, quantitative differences are observed for $T\gtrsim 80 \textmd{ MeV}$. While this bound is comparable to recent results in chiral perturbation theory to next-to-leading order~\cite{Adhikari:2020kdn}, where the agreement persists up to about $T\gtrsim 40 \textmd{ MeV}$, other models, such as the Nambu-Jona-Lasinio~\cite{He:2005nk} or the Polyakov loop-extended quark meson model~\cite{Adhikari:2018cea,Folkestad:2018psc}, for instance, show better agreement, both qualitatively and quantitatively, with the lattice phase diagram (see also Fig.~9 of Ref.~\cite{Mannarelli:2019hgn}). In particular, both models reproduce the steep rise in combination with the leveling-off of the BEC phase boundary at large $\mu_I$. Nonetheless, the qualitative agreement between the effective mass model and the lattice data together with the fact that the lattice results for $\mu_I^{\rm crit}$ and $\mu_Q^{\rm crit}$ do not differ by more than a few percents
again confirms that our model 
represents a reasonable approximation to the phase diagram at nonzero isospin (charge) densities. In addition, the inclusion of further hadrons and resonances in the effective mass model is straightforward.

\section{Primordial Black Holes Formation}
\label{app:pbhform}

\begin{figure}[t]
  \centering
  \includegraphics[width=.49\textwidth]{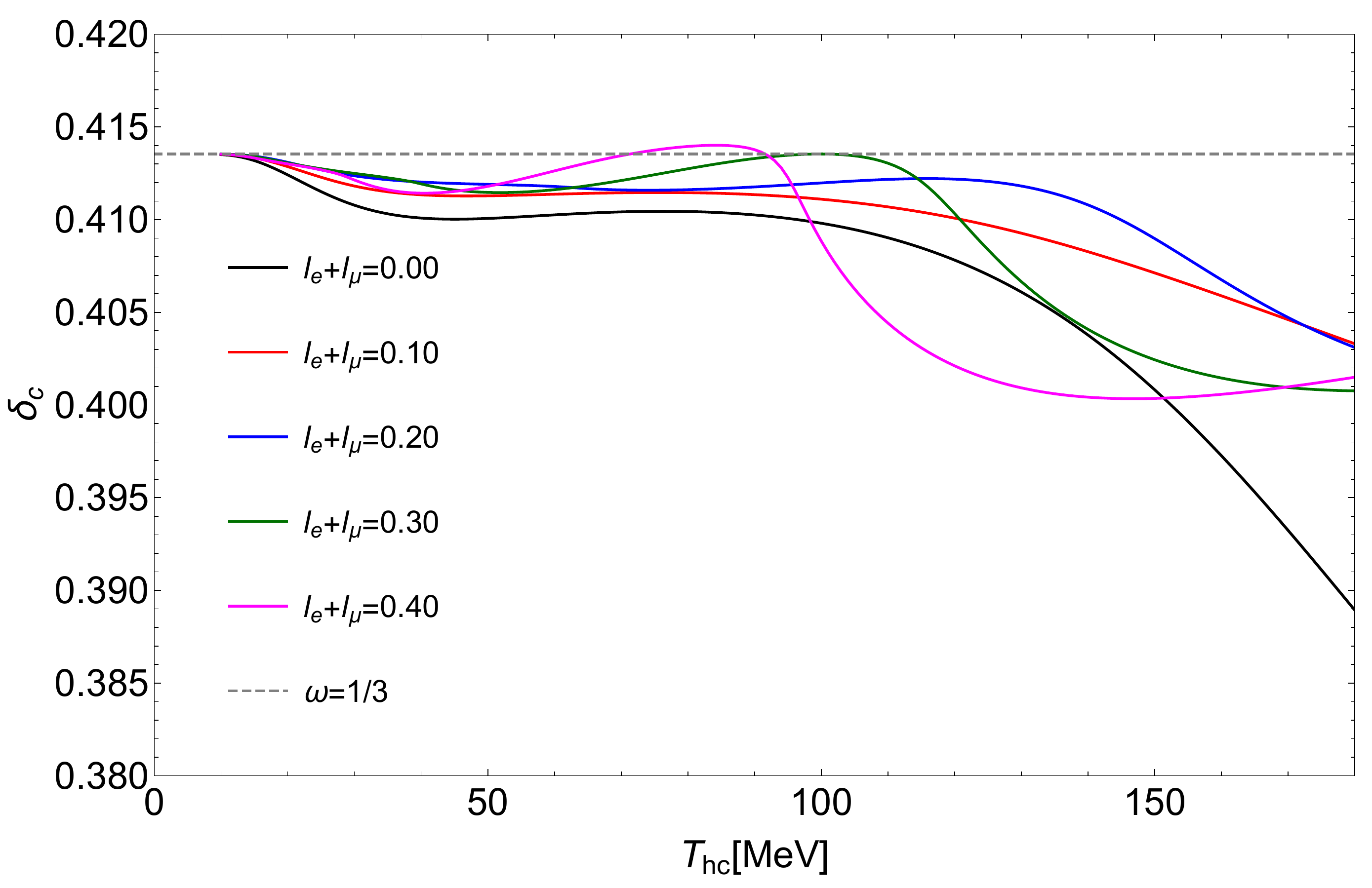}
  \caption{
  Threshold of primordial black hole formation versus horizon crossing temperature  for different values of the lepton asymmetry. 
  } \label{fig:pbh-th}
\end{figure}

At the time of PBH formation a region of the Universe within the Hubble horizon starts to collapse due to local inhomogeneities amounting to
\begin{eqnarray}
\beta(M)=\frac{M_{eq}}{M}\beta_{eq}\,.
\label{betamh}
\end{eqnarray}
The relation between the amplitude of the density perturbation $\delta$, the PBH mass $M_{\text{BH}}$ and the horizon mass $M_h$ can be defined as \cite{Niemeyer:1997mt,Niemeyer:1999ak}
\begin{eqnarray}
\label{threshold}
\delta=\left(\frac{M_{\text{BH}}}{K M_{h}}\right)^{\frac{1}{\gamma}}+\delta_c\,.
\end{eqnarray}
 The parameters in Eq. (\ref{threshold})  are obtained from numerical simulations to be  $K=3.3$ and $\gamma=0.36$ \cite{evans1994critical,Koike:1995jm,Niemeyer:1997mt,Niemeyer:1999ak,Musco:2008hv,Musco:2004ak}. The parameter $\delta_c$ is the threshold for PBH formation where different estimates for it exist in the literature \cite{Jedamzik:1996mr,Niemeyer:1997mt,Niemeyer:1999ak,Musco:2004ak,Musco:2008hv,Harada:2013epa,Escriva:2019phb}.  Here we assume that this threshold in a cosmological background slightly deviates from the one of a purely radiation dominated Universe ($\omega=p/\varepsilon=1/3$) which is estimated to be
$\delta_c\simeq 0.41$  \cite{Harada:2013epa,Escriva:2019phb}. Variation of $\delta_c$ due to different lepton asymmetry values is shown in Fig. \ref{fig:pbh-th}.

The fraction of PBHs  $\Omega_{\text{PBH}}$ with respect to the total cold dark matter (CDM) abundance $\Omega_{\text{CDM}}$ reads \cite{Byrnes:2018clq}
\begin{eqnarray}
\label{fracpbh}
f_{\text{PBH}}(M_{\text{\text{BH}}})=&&\frac{1}{\Omega_{\text{CDM}}}\int_0^\infty \frac{2 d M_{h}}{\sqrt{2\pi \sigma(M_h)^2}} \frac{M_{\text{BH}}}{\gamma M_{h}} \times
 \\ \nonumber &&\exp \left[-\frac{
 \delta^2(M_h)}{2\sigma^2(M_{h})}\right]\left(\frac{M_{\text{BH}}}{K M_{h}}\right)^{\frac{1}{\gamma}} \sqrt{\frac{M_{eq}}{M_{h}}}
\,.
\end{eqnarray}

The mass or scale dependence of the density perturbation width can be assumed to be  \cite{Byrnes:2018clq}
\begin{eqnarray}
\label{sigwid}
\sigma^2(M_h)=0.003\left(\frac{M_{h}}{10 M_{\odot}}\right)^{n_M}\,.
\end{eqnarray}
The density spectral index $n_M$ can be  related to the  scalar spectral index  $n_{S}-1\simeq -2 n_{M}$, where $n_S\simeq0.96$  \cite{Akrami:2018odb,Aghanim:2018eyx}. We choose the 
benchmark value of $n_M=0$ 
  to compute the fraction of PBH from Eq.~(\ref{fracpbh}). The parameter $f_{\text{PBH}}$ for masses smaller than $M_{\odot}$ increases (decreases) when $n_M$ is negative (positive). However, $f_{\text{PBH}}$ increases (decreases) for larger masses, respectively. For a fixed $n_M$ as lepton asymmetry increases the value of $f_{\text{PBH}}$ will change depending on the behavior of $\omega$ or  $\delta_c$ and the energy and pressure density.  When $\delta_c$ increases (decreases) the fraction of PBH decreases (increases).

\end{document}